\documentclass[final,5p,times,twocolumn]{elsarticle}
\usepackage{graphics, graphicx, float, amssymb}

\usepackage{graphicx}
\usepackage{color}
\usepackage{amssymb}
\journal{Journal of Physics and Chemistry of Solids}

\begin{document}

\begin{frontmatter}

\title{Inelastic x-ray scattering investigations of lattice dynamics in SmFeAsO$_{1-x}$F$_y$ superconductors}

\author[1]{Mathieu Le Tacon}\ead{m.letacon@fkf.mpg.de}%\corref{cor1}
\author[2,3]{T. R. Forrest}
\author[2]{Ch. R\"{u}egg}
\author[4]{A. Bosak}
\author[3,6]{J. Noffsinger}
\author[4,2]{A. C. Walters}
\author[7]{P. Toulemonde}
\author[8,9] {A. Palenzona}
\author[10]{N. D. Zhigadlo}
\author[10]{J. Karpinski}
\author[11]{J. P. Hill}
\author[4]{M. Krisch}
\author[2]{D.F. McMorrow}

\address[1]{Max-Planck-Institut f\"{u}r Festk\"{o}rperforschung; Heisenbergstra{\ss}e 1; D-70569 Stuttgart, Germany}
\address[2]{London Centre for Nanotechnology and Department of Physics and Astronomy; University College London; London WC1E 6BT; United Kingdom}
\address[3]{Department of Physics, University of California at Berkeley, Berkeley, California 94720}
\address[4]{European Synchrotron Radiation Facility; BP 220; F-38043 Grenoble Cedex; France}
\address[6]{Materials Sciences Division, Lawrence Berkeley National Laboratory, Berkeley, California 94720, USA}
\address[7]{Institut N\'{e}el, CNRS/UJF; Grenoble - France}
\address[8]{CNR-INFM-LAMIA Artificial and Innovative Materials Laboratory, Corso Perrone 24, 16152 Genova, Italy}
\address[9]{Department of Chemistry and Industrial Chemistry, University of Genova, via Dodecaneso 31, 16146 Genova, Italy}
\address[10]{Laboratory for Solid State Physics ETH Z\"{u}rich; CH-8093 Z\"{u}rich; Switzerland}
\address[11]{Department of Condensed Matter Physics and Materials Science; Brookhaven National Laboratory; Upton; New York 11973, USA}

\cortext[cor1]{Corresponding author.}

%$^{4}$Solid State Physics Division; Bhabha Atomic Research Centre; Trombay; Mumbai 400 085; India\\
%$^{5}$Laboratory for Quantum Magnetism; Ecole Polytechnique F\'ed\'erale de Lausanne; CH--1015 Lausanne; Switzerland\\
%$^{6}$Laboratory for Solid State Physics ETH Z\"{u}rich; CH-8093 Z\"{u}rich; Switzerland\\
%$^{7}$Department of Condensed Matter Physics and Materials Science; Brookhaven National Laboratory; Upton; New York 11973, USA\\}

\begin{abstract}
%A self-contained abstract not exceeding 200 words outlining in a single paragraph the aims, scope and conclusions of the paper must be supplied.
We report measurements of the phonon density of states as measured with inelastic x-ray scattering in SmFeAsO$_{1-x}$F$_y$ powders. An unexpected strong renormalization of phonon branches around 23 meV is observed as fluorine is substituted for oxygen. Phonon dispersion measurements on SmFeAsO$_{1-x}$F$_y$ single crystals allow us to identify the 21 meV A$_{1g }$ in-phase (Sm,As) and the 26 meV B$_{1g }$ (Fe,O) modes to be responsible for this renormalization, and may reveal unusual electron-phonon coupling through the spin channel in iron-based superconductors.
\end{abstract}


\end{frontmatter}

%% \linenumbers

%% main text

\section{Introduction}
\label{}

The recent discovery of superconductivity in FeAs based compounds has sparked a new gold rush amongst the strongly correlated electron community~\cite{Takahashi_Nature2008}. Among the various families of FeAs based superconductors, the so-called '1111', with composition  REFeAsO (RE being a rare earth), has the highest values of the superconducting transition temperatures $T_c$: depending on the lanthanide, $T_c$ can be as high as 55 K upon fluorine doping~\cite{Ren_CPL2008}. The high value of $T_c$, early DFT calculations~\cite{Boeri}, and other physical properties, appeared to support the notion of an exotic mechanism leading superconductivity in these materials. There is however increasing speculation that an exotic electron-phonon coupling, possibly enhanced through the spin or orbital channels, is at play in the iron-based high-temperature superconductors. In particular, the fact that the Fe-As configuration is intimately related to $T_c$~\cite{Mukuda_JPSJ2008} and to the magnetic state of the iron sublattice~\cite{Yin_PRL2008, Yildirim_PhysicaC2009} show the lattice is active in this problem. This was further confirmed recently through the observation of strong renormalization of zone center Raman~\cite{Rahlenbeck,Chauviere_PRB2009} or infrared~\cite{Akrap_PRB2009} active phonons at $T_N$ in BaFe$_2$As$_2$.
Here, we present inelastic x-ray scattering (IXS) measurements on poly- and single-crystalline samples of SmFeAsO$_{1-x}$F$_y$ that allowed us to investigate the doping dependence of the generalized phonon density of states (GPDOS) and of the phonon dispersions. We report an unexpectedly strong doping-induced renormalization of two phonon branches - the 21 meV (As,Sm) and 26 meV (Fe,O) modes - providing significant insight into the evolution with doping of the energy and momentum dependence of the coupling between the lattice and the spin degrees of freedom, and suggesting that the spin-phonon coupling may be of key importance in these compounds.

\section{Results}

\par
SmFeAs(O$_{0.9}$F$_{0.1}$) samples (Sm-1111) was prepared under high pressure and high temperature using a "belt" type high pressure cell in a similar way to that reported previously for LaFeAsO$_{1-x}$F$_x$~\cite{Garbarino_PRB2008} . Sm, Fe, Fe$_2$O$_3$, As and SmF$_3$ (in the case of the fluorine doped sample) powders were mixed together and pressed in the form of cylindrical pellets. They were then introduced into a crucible (machined from rods of hexagonal boron nitride) surrounded by a cylindrical graphite resistive heater and the whole assembly was placed in a pyrophyllite gasket. The samples were treated at 6GPa, 1000-1100$\,^{\circ}{\rm C}$ for 4 hours, then quenched to room temperature and depressurized. The XRD patterns show that the major phase is the one expected, Sm-1111 with some small impurities of FeAs, SmAs and Sm$_2$O$_3$. SmFeAsO synthesis is described in ref.~\cite{Martinelli_SSC2008}.

\begin{figure}%[ptbh]
\begin{center}
\includegraphics[width=0.7\columnwidth]{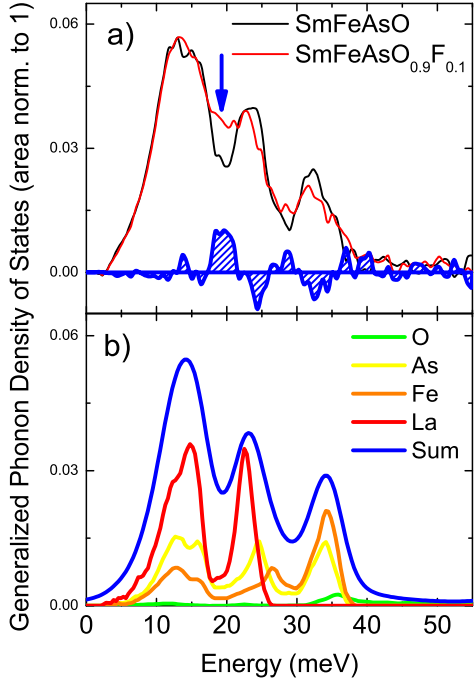}
\end{center}\vspace{-5mm}
\caption{a) comparison between the measured generalized phonon density  of states of SmFeAsO (black curve) and SmFeAsO$_{0.9}$F$_{0.1}$ (red curve), measured at room temperature. The area of both spectra have been normalized to unity. The blue line corresponds to the difference between the two spectra. b) partial PDOS obtained from DFT calculation~\cite{Boeri} and weighted according to the procedure described in ref. ~\cite{LeTacon_PRB2008}.}
\label{fig1}
\end{figure}

\par
The IXS experiment on polycrytalline SmFeAsO and SmFeAsO$_{0.9}$F$_{0.1}$ was carried out on beamline ID28 at the ESRF using the silicon (9 9 9) configuration at 17.794 keV with an instrumentation energy resolution of 3.0 meV.
Within the limit of single phonon scattering (Stokes process), the intensity of the inelastically scattered signal with energy (momentum) transfer $\omega$ ($\textbf{Q}=(Q_a~Q_b~Q_c)$) is directly proportional to the dynamical structure factor~\cite{Burkel_RPP99}:

\begin{eqnarray}
\label{SofQ}
\nonumber
S(\textbf{Q},\omega) & =& \sum_j \bigg | \sum_n \frac{f_n(\textbf{Q})e^{i\textbf{Q}.\textbf{r$_n$} - W_n}(\textbf{Q}.\hat \sigma_n(\textbf{Q}, j))}{\sqrt{M_n}}\bigg |^2\\
&  & \times \frac{(1+n(\omega_{\textbf{Q}, j},T))\delta(\omega-\omega_{\textbf{Q}, j})}{\omega_{\textbf{Q}, j}}\\
\nonumber
\end{eqnarray}
where $f_n(\textbf{Q})$ is the atomic form factor of the $n^{th}$ atom (at position $\bf{r_n}$) in the unit cell, $M_n$ its mass and $W_n$ the Debye-Waller factor. The summation index $j$ refers to the $j^{th}$ phonon branch and $\hat \sigma_n(\textbf{Q}, j)$ and $\omega_{\textbf{Q}, j}$ to the associated eigenvector and eigenvalues,  respectively. Finally, $n(\omega, T)=(exp(\omega/k_B T)-1)^{-1}$ stands for the Bose factor.
The scattered photons are analyzed by a set of 8 analyzers that were centered on two different set of angles to cover a $Q$ range from 53 nm$^{-1}$ to 73 nm$^{-1}$ ($\langle Q \rangle =$ 63 $nm^{-1}$), and in this case the dynamical structure factor must be averaged over the sphere of radius $Q = \|\textbf{Q}\|$. As previously demonstrated~\cite{Bosak_PRB2005}, averaging several measurements over a large range of $Q$ ($Q_{min} < Q < Q_{max})$ gives access to:
\begin{eqnarray}
\label{SofQ2}
S(\omega) & \propto & \frac{(1+n(\omega,T))}{\omega} \times GPDOS(\omega),\\
\nonumber
\end{eqnarray}

with the GPDOS being defined as a weighted sum of the partial phonon densities of states, $G_n(\omega)$, for each atomic species

\begin{eqnarray}
\label{SofQ3}
GPDOS(\omega) & = & \sum_n\frac{e^{-2W_n} f_n(\langle Q \rangle)^2 G_n(\omega)}{M_n}\\
\nonumber
\end{eqnarray}

In the upper panel of Figure~\ref{fig1}, we show the quantity $I_{inelastic}(\omega)\times [\omega/(1+n(\omega,T))]$ measured at room temperature on SmFeAsO and nominally doped SmFeAsO$_{0.9}$F$_{0.1}$. The inelastic signal $I_{inelastic}(\omega)$ has been obtained after summation of the individual spectra weighted by the appropriate analyzer efficiency, subtraction of the elastic contribution, and recursive correction from multiphonon contributions, following refs.~\cite{Bosak_PRB2005, Kohn_HI2000}.
As previously reported for the case of NdFeAsO$_{1-x}$F$_x$~\cite{LeTacon_PRB2008} and LaFeAsO$_{1-x}$F$_x$~\cite{Fukuda_JPCS2008}, we observe essentially three peaks of about 5-6 meV full-width-half-maximum located around 13, 24 and 32 meV.
In the lower panel of Figure~\ref{fig1}, we report the theoretical results obtained by Boeri et al.~\cite{Boeri} on a similar compound, LaOFeAs.
To allow a direct comparison between the experimental data and the theoretical calculation, we have weighted each calculated partial PDOS $G_n(\omega)$ by its corresponding $e^{-2W_n} f_n(\langle Q \rangle)^2 M_n^{-1}$ factor (these factors are given in ref.~\cite{LeTacon_PRB2008}). The result of this procedure has been plotted in the lower panel of Fig.~\ref{fig1}.
The overall agreement with the experimental data for the parent compound is good for the two first peaks, but as previously observed in the cases of NdFeAsO and SmFeAsO, the higher energy peak is found to be harder, by about 4 meV, calculated than is observed in the experiment.

The GPDOS measured on SmFeAsO$_{0.9}$F$_{0.1}$ and plotted in the upper panel of Figure~\ref{fig1} reveal a spectra with a similar overall shape to the one measured on SmFeAsO, except for the second peak. There is indeed, a striking change below 24 meV, with a clear increase of the spectral weight around 21 meV similar to our previous findings in NdFeAsO$_{1-x}$F$_x$~\cite{LeTacon_PRB2008}.

According to first principle calculations, this middle band originates from three optical phonons branches extending from 20 to 28 meV. These modes are Raman active at the zone center and are associated with the in-phase A$_{1g}$ (As,Sm) vibration ($\sim$ 21 meV at $\bf{Q}=0$ in SmFeAsO from Raman measurements~\cite{Hadjiev_PRB2008, Marini_EPL2008}), the out-of-phase A$_{1g}$ (As,Sm) mode (A$_{1g}$(2) - 23 meV) and  the B$_{1g}$ (Fe,O) mode (26 meV), respectively.

To go further and determine which of these modes are responsible for the observed renormalization, we have carried out IXS experiments on SmFeAsO and superconducting SmFeAsO$_{0.6}$F$_{0.35}$ single crystals to measure their phonon dispersion, with particular emphasis on these three modes.
High quality single crystals of SmFeAsO and SmFeAsO$_{0.60}$F$_{0.35}$ of about 100$\times$100$\times$20 $\mu $m$^3$ were grown as described elsewhere~\cite{Zhigadlo_JPCM2008, Karpinski_PhysicaC2009}.
Our SQUID measurements on SmFeAsO show magnetic ordering at $T_N =$ 130~K, and the superconducting transition in the nominally doped SmFeAsO$_{0.60}$F$_{0.35}$ takes place at $T_c =$ 53~K.
We used an incident photon energy of 21.747 keV (silicon (11 11 11) configuration) providing an energy resolution of 2.0 meV on ESRF beamline ID28. The x-ray beam was focused down to 50$\times$40 $\mu m^2$. Rocking curves of 0.5$^\circ$ were measured on the (0 0 4) reflection, indicating a good c-axis mosaicity.

From eq.~\ref{SofQ}, one can see that probing c-axis polarized phonons, such as of those we are interested in, requires one to carry out measurements in a zone with the largest possible $Q_c$ component along the $c^*$ reciprocal lattice vector (note that in what follows, $\textbf{Q}$ is defined with respect to the tetragonal unit cell, and $\textbf{q}=(q_a~q_b~q_c)$ refers to the reduced momentum transfer relative to the zone center).
To enhance the contrast of the optical modes with respect to the strong acoustic one, we also decided to work close to a weak Bragg spot. The best compromise was found to be the zone around $\Gamma$ = (1 0 12). This corresponds to a challenging quasi-grazing incidence reflection geometry (because of the size of the crystals), with a scattering angle $2\Theta \sim$ 49$^\circ$. This angle is close to the highest achievable for ID28, and we had to perform the transverse scans along the $\overline{1}00$ direction.

\begin{figure}%[ptbh]
\begin{center}
\includegraphics[width=\columnwidth]{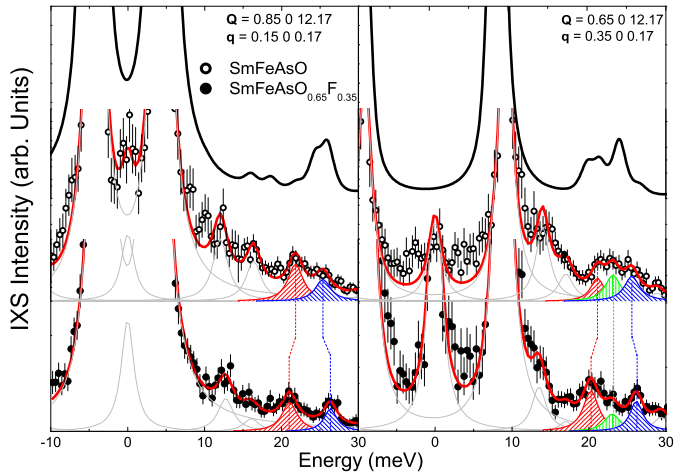}
\end{center}\vspace{-5mm}
\caption{Experimental IXS spectra of SmFeAsO$_{1-x}$F$_y$ for \textbf{q} =(0.15~0~0.15) (left) and \textbf{q} =(0.35~0~0.17) (right) measured in the Brillouin zones centered around $\Gamma$ = (1~0~12). The red lines are the result of the total fit, and the thin gray lines the individual phonons. The in-phase (red) and out-of-phase (green) A$_{1g}$ (As,Sm) modes as well as the B$_{1g}$ (Fe,O) mode (blue) have been highlighted. The black lines represent the theoretical spectra calculated for LaFeAsO at these \textbf{Q} vectors from the DFT ~\cite{Noffsinger_PRL2009}.}
\label{figexpvsth}
\end{figure}

In Fig.~\ref{figexpvsth} we present typical IXS spectra obtained for SmFeAsO and SmFeAsO$_{0.65}$F$_{0.35}$, together with the result from DFT calculation performed for LaFeAsO from ref.~\cite{Noffsinger_PRL2009}.
Experimental data were analyzed by fitting to a series of Lorentzians, as shown in Fig.~\ref{figexpvsth}. With the exception of the elastic line, fitting of the data to resolution limited peaks gave rather poor results, and slightly broader lineshapes (2.2 to 2.5 meV) were used. Similar broadening has also been reported in the case of CaFe$_2$As$_2$~\cite{Mittal_PRL2009}.
The assignment of the phonon branches is challenging due to the large discrepancies between the IXS intensities predicted by the DFT calculations and the experimental spectra, as illustrated in Fig.~\ref{figexpvsth}. We by-passed this difficulty, at least for the three c-axis polarized branches around 23 meV we are interested in, by taking advantage of measurements performed in an adjacent Brillouin zone (centered around $\Gamma$ = 1 0 11), as described in ref.~\cite{Letacon_PRB2009}.

\begin{figure}%[!t]
\begin{center}
\includegraphics[width=1\columnwidth]{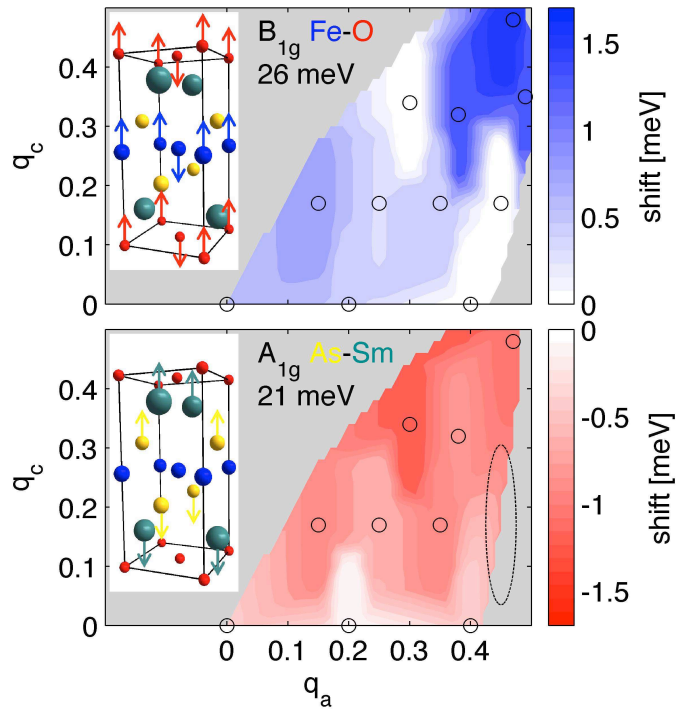}
\end{center}\vspace{-5mm}
\caption{Momentum dependence of the doping induced renormalization of the 26 meV,  B$_{1g}$ (upper panel) and 21 meV, in-phase A$_{1g}$ (lower panel) modes in the ($q_a$~0~$q_c$) plane. The empty circles correspond to the analyzer positions, the dashed ellipse indicates the q-resolution, and the white areas where there is insufficient data. The insets show schematics of the respective \textbf{q}=0 eigenvectors, with the atoms color coded as indicated (from ref.~\cite{Letacon_PRB2009}).}
\label{figmaps}
\end{figure}

Using this method and exploiting the 8 analyzers available on ID28, we have been able to map out the dispersion of the three c-axis polarized modes in the ($q_a~0~q_c$) plane for the parent and doped compounds, at room temperature~\cite{Letacon_PRB2009}.
The renormalization of the modes was found to be rather weak in the $\Gamma X$ direction, but displayed a surprisingly strong dependence with $q_c$. As shown in Fig.~\ref{figexpvsth}, we observed a softening of the in-phase A$_{1g}$ mode and a hardening of the B$_{1g}$ mode. The maximum amplitude for these effects are obtained close to \textbf{q}=(0.3 0 0.3) for the in-phase A$_{1g}$ mode (1.2 meV softening) and close to \textbf{q}=(0.5 0 0.5) for the B$_{1g}$ mode (1.7 meV hardening). No clear doping dependence for the out-of-phase A$_{1g}$ (As,Sm) mode was found.

\section{Discussion}

The disagreement between the IXS intensities predicted by standard DFT calculation and experimental data has also been reported in the case of the 122 compounds~\cite{Mittal_PRL2009, Hahn_PRB2009, Reznik_PRB2009}.
In these cases a much better agreement between theory and experiment in parent compounds above their magnetic transition has been found by including spin-polarization in the calculations. Ref.~\cite{Reznik_PRB2009} also mentions the fact that it improves the agreement between the experimental and calculated GPDOS in the case of the 1111 compounds.
This indicates that magnetism is the missing ingredient in the {\em ab-initio} calculations discussed above.
It was also found that the magnetic ground state was intimately related to the Fe-As distance along the c-axis~\cite{Yin_PRL2008, Yildirim_PhysicaC2009}, which would naturally lead to a coupling between the atomic motion modulating this distance - precisely the in-phase out-of-plane A$_{1g}$ (As,Sm) and the (Fe,O) B$_{1g}$ modes - and the amplitude of the Fe moments~\cite{Yndurain_PRB2009}.

As far as the doping dependence of the phonon spectra is concerned, a calculation based on the virtual crystal approximation suggested that doping should induce only minor changes in the phonon spectra~\cite{Boeri}. Another study taking into account the F-doping explicitly in a 2x2x1 supercell within an LDA approach~\cite{Noffsinger_PRL2009} has related the doping-induced changes seen in the PDOS~\cite{LeTacon_PRB2008} to structural relaxation (with a relative change of the unit cell volume $\Delta V/V$ of about 4 $\%$). However,  this approach leads to a hardening of the in-phase (As,Sm) A$_{1g}$ mode that contrasts with the observed softening, and overestimates the structural changes as experimentally, the doping-induced unit cell volume is rather limited ($\Delta V/V \sim 0.5 \%$).
The observed doping-induced phonon renormalization and their momentum dependencies reported here may rather reflect the evolution upon fluorine doping of momentum and frequency of either the electron-phonon coupling constant $g(\textbf{q},\omega)$ or the electronic/magnetic susceptibility $\chi(\textbf{q},\omega)$ of the system.
To fully understand the role played by the coupling of phonon to magnetism in iron-pnictides, further theoretical and experimental investigations are needed. We also note that in a recent calculation for the case of 122 compounds, it was shown that, in addition to a renormalization of the phonon frequencies, magnetism also tends to enhance the electron-phonon coupling constant, though this does not appear to be sufficient to explain the high superconducting $T_c$ of these materials~\cite{Boeri_2010}.

\section{Conclusion}

In summary, we have found a clear doping dependence for two c-axis polarized phonon branches in the 1111 iron-pnictide family that may reveal unusual electron-phonon coupling through the spin channel in iron-based superconductors. Further theoretical and experimental investigations are needed to fully understand the role played by such coupling in these materials properties.

\section{Aknowledgements}
This project was supported by the U.S. Department of Energy, Division of Materials Science, under Contract No. DE-AC02-98CH10886,  by the Swiss National Science Foundation (NCCR MaNEP), ANR SupraTetraFer, the Royal Society, and EPSRC.

\end{document}